\documentstyle[12pt]{article}
\title{A modification of the 10d superparticle action\\
inspired by the Gupta-Bleuler quantization scheme}
\author{S. Bellucci\thanks{e-mail: bellucci@lnf.infn.it}\\
INFN-Laboratori Nazionali di Frascati, \\
P.O. Box 13, 00044 Frascati, Italy\\
 and \\
A. Galajinsky\thanks{e-mail: galajin@fma.if.usp.br, permanent address:
Department of Theoretical Physics,
Tomsk State University, 634050 Tomsk,
Russian Federation}\\
Instituto de Fisica, Universidade de S\~ao Paulo, \\
C. Postal 66318, 05315-970, S\~ao Paulo, Brasil}
\date{}
\begin{document}
\maketitle
\large
\begin{abstract}
We reconsider the issue of the existence of a complex
structure in the Gupta-Bleuler quantization scheme. We prove an
existence theorem for the complex structure associated with the $d=10$
Casalbuoni-Brink-Schwarz superparticle, based on an explicitly
constructed Lagrangian that allows a holomorphic-antiholomorphic
splitting of the fermionic constraints consistent with the vanishing
of all first class constraints on the physical states.
\end{abstract}

\vspace{0.5cm}

As it is well known, the puzzle of the covariant quantization of the
superparticle, superstring models can be viewed as the problem of
mixed first and second class fermionic constraints in the Hamiltonian 
formalism [1--3]. One of the interesting approaches to treat the
second class constraints is the Gupta-Bleuler-type quantization scheme
[4-7] which, for the case at hand, reduces to the construction of a
specific complex structure $J$ on a phase space of the models 
\footnote{For simplicity, in what follows we shall discuss the
superparticle case only. A discussion of questions related to the
covariant quantization of the Green-Schwarz superstring can be found
e.g. in [8,9]}. The latter provides a
holomorphic-antiholomorphic splitting of the mixed constraints which
proved to yield a successful covariant quantization of the $4d$ 
superparticle [5].

A recipe how to construct such a $J$ in arbitrary space-time
dimensions has been proposed in the recent work [10]. The strategy
adopted then was to decompose the tensor $J$ into irreducible
representations (irreps) of the Lorentz group and then reduce the
equations for 
determining $J$ to those for the irreps. The explicit solution in $d=10$
has been found [10]
\addtocounter{equation}{1}
$$J_{ab}=\frac 1\alpha (A_aB_b-A_bB_a),
\eqno{(1a)}$$
$$ \alpha=\pm\sqrt{A^2B^2-(AB)^2},
\eqno{(1b)}$$
$$ (Ap)=0, \qquad (Bp)=0,
\eqno{(1c)}$$
requiring the extension of the original phase space $(x^n,p_n),
(\theta^{\alpha},p_{\theta\alpha})$ through the new vector 
variables $A^n,B^n$. Generally, such an extension can easily be realized by
introducing two pairs of canonically conjugate variables
$(A^n,{p_A}_n)$,$(B^n,{p_B}_n)$ subject to the first class constraints
\begin{equation}
{p_A}_n=0, \qquad {p_B}_n=0,
\end{equation}
and treating the equations (1c) as gauge fixing conditions for some
of the constraints (2). However, as the first class constraints 
remaining in Eq.(2) do not commute with the complex structure, in
passing to a quantum description the vanishing of these constraints on
physical states would be incompatible with the vanishing of the
holomorphic constraints on those states.

In this brief note we suggest a way to cure this inconsistency.
The idea is to completely fix the
gauge freedom in the sector $(A,p_A)$, $(B,p_B)$ by introducing further
auxiliary variables. If the first class constraints from the sector of
the new variables turn out to commute with $A$ and $B$, the complete
description is self-consistent.

The action to be examined reads
\begin{eqnarray}
&& S=\displaystyle\int d\tau \frac 1{2e} (\dot x^n-i\theta\Gamma^n
\dot\theta-\omega_1A^n-\omega_2B^n-\mu_i{\Lambda^n}_i)^2 -
\rho_1(A^2-1)-\cr
&&-\rho_2(B^2-1)-\nu_{1i}(A\Lambda_i)-\nu_{2i}(B\Lambda_i)-
\Phi_{ij}(\Lambda_i\Lambda_j+\Delta_{ij})-\sum_{i=1}^{8}\mu_i
\end{eqnarray}
where
\[ \Delta_{ij}\equiv\left\{
\begin{array}{rl}
0, & \mbox i=j\\
1, & \mbox i \ne j\\
\end{array} \right.   i,j=1,\dots,8 \]
Here the summation over repeated indices is understood. As compared to the
Casalbuoni-Brink--Schwarz model [11] one finds a set of auxiliary variables
($A^n,B^n,{\Lambda^n}_i,\mu_i,{\nu_1}_i,
{\nu_2}_i,\Phi_{ij},\omega_1,\omega_2,\rho_1,\rho_2$), with $\Phi_{ij}$
being symmetric.

Consider the model (3) in the Hamiltonian formalism. Introducing
momenta ($p_e,p^n,p_{\theta\alpha},{p_A}^n,{p_B}^n,{p_\Lambda}_{ni},
{p_\mu}_i,{p_{\nu_1}}_i,{p_{\nu_2}}_i,{p_\Phi}_{ij},p_{\omega_1},p_{\omega_2},
p_{\rho_1},p_{\rho_2}$) canonically conjugate to the configuration
space variables one has a set of primary constraints
\addtocounter{equation}{1}
$$ p_e=0, \qquad p_\theta+i\theta\Gamma^np_n=0,\\
\eqno{(4a)}$$
$$ {p_A}^n=0, \qquad {p_B}^n=0, \qquad {p_\Lambda}_{ni}=0,\\
\eqno{(4b)}$$
$$ {p_\mu}_i=0, \qquad {p_{\nu_1}}_i=0, \qquad{p_{\nu_2}}_i=0,\\
\eqno{(4c)}$$
$$ p_{\omega_1}=0,\qquad p_{\omega_2}=0,\qquad p_{\rho_1}=0,\\
\eqno{(4d)}$$
$$ p_{\rho_2}=0, \qquad p_{\Phi{ij}}=0,
\eqno{(4e)}$$
and the relation to eliminate $\dot x^n$
\begin{equation}
\dot x_n=ep_n+i\theta\Gamma_n\dot\theta+\omega_1A_n+\omega_2B_n+
\mu_i{\Lambda_n}_i.
\end{equation}
The canonical Hamiltonian is
\begin{eqnarray}
&& H=(p_\theta+i\theta\Gamma^np_n)\lambda_\theta+p_e\lambda_e+
p_A\lambda_A+p_B\lambda_B+{p_\Lambda}_i{\lambda_\Lambda}_i
+{p_\mu}_i{\lambda_\mu}_i+\cr
&&+{p_{\nu_1}}_i{\lambda_{\nu_1}}_i+{p_{\nu_2}}_i{\lambda_{\nu_2}}_i+
p_{\Phi_{ij}}\lambda_{\Phi_{ij}}+p_{\omega_1}\lambda_{\omega_1}+p_{\omega_2}
\lambda_{\omega_2}+
p_{\rho_1}\lambda_{\rho_1}+p_{\rho_2}\lambda_{\rho_2}+\cr
&&+e\displaystyle\frac{p^2}2+\omega_1(pA)+\omega_2(pB)+\rho_1(A^2-1)+
\rho_2(B^2-1)+\nu_{1i}(A\Lambda_i)+\cr
&&+\nu_{2i}(B\Lambda_i)+\Phi_{ij}(\Lambda_i\Lambda_j+\Delta_{ij})+
\mu_1((p\Lambda_1)+1)+\mu_2((p\Lambda_2)+1)+\dots\cr
&&+\mu_8((p\Lambda_8)+1),
\end{eqnarray}
where the $\lambda's$ denote Lagrange multipliers corresponding to the
primary constraints.

The consistency conditions for the primary constraints imply the secondary
ones \footnote {We define the Poisson brackets of the variables
($\Lambda,p_ {\Lambda}),(\Phi,p_{\Phi})$ in the form
$\left \{{\Lambda^n}_i,{p_\Lambda}_{mj} \right\}={\delta^n}_m\delta_{ij},$
$\left \{\Phi_{ij},p_{\Phi ks} \right \}=\frac 12 \left
(\delta_{ik}\delta_{js}+ \delta_{is}\delta_{jk} \right ).$}
\addtocounter{equation}{1}
$$
p^2=0,\qquad p\Lambda_i+1=0, \qquad
\Lambda_i\Lambda_j+\Delta_{ij}=0,
\eqno{(7a)}$$
$$ pA=0, \qquad A^2-1=0, \qquad A\Lambda_i=0,
\eqno{(7b)}$$
$$ pB=0, \qquad B^2-1=0, \qquad B\Lambda_i=0,
\eqno{(7c)}$$
$$ \omega_1 p^n+2\rho_1 A^n+\nu_{1i}{\Lambda^n}_i=0,
\eqno{(7d)}$$
$$ \omega_2 p^n+2\rho_2 B^n+\nu_{2i}{\Lambda^n}_i=0,
\eqno{(7e)}$$
$$ \nu_{1i} A^n+\nu_{2i}B^n+\mu_i p^n+2\Phi_{ij}{\Lambda^n}_j=0,
\eqno{(7f)}$$
and determine half of the $\lambda_\theta$
\begin{equation}
\Gamma^np_n\lambda_\theta=0.
\end{equation}
Consider now Eq. (7d). Multiplying it by $A^n$ and taking into account
Eq. (7b) one gets
\begin{equation}
\rho_1=0.
\end{equation}
Subsequent multiplication of the remaining equation $\omega_1 p^n+
\nu_{1i}{\Lambda^n}_i=0$ by $p^n,{\Lambda^n}_i$ reduces it to a system of
linear homogeneous equations which has the trivial solution
\begin{equation}
\omega_1=0, \qquad \nu_{1i}=0,
\end{equation}
since the
matrix
$$
\begin{array}{l}
\left(\begin{array}{cc}
p^2 & p\Lambda_i\\
p\Lambda_j & \Lambda_i\Lambda_j
\end{array}\right),
\end{array}$$
is nondegenerate on the constraint surface (7a)-(7f). In the same
spirit Eqs. (7e),(7f) simplify to
\begin{equation}
\omega_2=0, \qquad \rho_2=0,\qquad \nu_{2i}=0, \qquad \mu_i=0, \qquad
\Phi_{ij}=0.
\end{equation}
The preservation in time of the secondary
constraints (7a)-(7c),(9)-(11) determine some of the Lagrange
multipliers
\addtocounter{equation}{1}
$$ p\lambda_A=0,\qquad  A\lambda_A=0,\qquad
\Lambda_i\lambda_A+A\lambda_{\Lambda i}=0\\
\eqno{(12a)}$$
$$ p\lambda_B=0,\qquad  B\lambda_B=0,\qquad
\Lambda_i\lambda_B+B\lambda_{\Lambda i}=0\\
\eqno{(12b)}$$
$$ p\lambda_{\Lambda i}=0,\qquad  \Lambda_i\lambda_{\Lambda j}+
\Lambda_j\lambda_{\Lambda i}=0,\\
\eqno{(12c)}$$
$$
\lambda_{\rho_1}=0, \qquad  \lambda_{\omega_1}=0, \qquad
{\lambda_{\nu_1}}_i=0,\\
\eqno{(12d)}$$
$$ \lambda_{\rho_2}=0, \qquad  \lambda_{\omega_2}=0, \qquad
{\lambda_{\nu_2}}_i=0,\\
\eqno{(12e)}$$
$$ \lambda_{\mu i}=0, \qquad  \lambda_{\Phi ij}=0,
\eqno{(12f)}$$
and no tertiary constraints appear.

Taking into account Eqs. (4),(9)-(11) one concludes that the
variables
$(\rho_1,p_{\rho_1}),(\rho_2,p_{\rho_2}),(\mu_i,{p_\mu}_i),
({\nu_1}_i,{p_{\nu_1}}_i),({\nu_2}_i,{p_{\nu_2}}_i),(\Phi_{ij},{p_\Phi}_{ij}),
(\omega_1,p_{\omega_1}),\\
(\omega_2,p_{\omega_2})$ are unphysical and can be omitted after introducing
the associated Dirac bracket.  Thus, the only nontrivial constraints to be
analyzed are those from Eqs.  (7a)-(7c), together with the corresponding
momenta (4b).

Let us now return to Eq. (7). The constraints (7b) together with the
corresponding momentum $p_{An}=0$ are second class. In a full agreement
with this, Eq. (12a) involving the associated Lagrange multiplier
$\lambda_A$ can be solved explicitly. Actually, since the vectors
$p^n,A^n,{\Lambda^n}_i$ satisfying Eq. (7) are linearly independent for any
fixed value compatible with Eq. (7), the matrix
\begin{equation}
\begin{array}{lll}
\left(\begin{array}{ccc}
p_0 & \dots & p_9 \cr
A_0 & \dots & A_9 \cr
\Lambda_{0i} & \dots & \Lambda_{9i} \cr
\end{array}\right),
\end{array}
\end{equation}
is invertible on the constraint surface. The latter fact implies that the
system of linear inhomogeneous equations (12a) has a unique solution for
any fixed value of $p^n,A^n,{\Lambda^n}_i$. Analogously, the constraints
(7c) and $p_{Bn}=0$ are second class and Eq. (12b) uniquely determines
$\lambda_B$.

Thus, it remains to discuss the constraints (7a) and the corresponding
momentum $p_{\Lambda ni}=0$. In order to extract the first class constraints
contained in $p_{\Lambda ni}=0$, it suffices to construct operators
projecting onto subspaces orthogonal to $(A^n,B^n)$ and
$(p^n,\Lambda_{ni})$ respectively.  The explicit form of the projectors
is
\begin{eqnarray}
&& {{\Pi_{(A,B)}}^m}_n={\delta^m}_n+\frac {(AB)}{1-(AB)^2} A^mB_n+
\frac {(AB)}{1-(AB)^2} B^mA_n- \cr
&&-\frac {1}{1-(AB)^2} A^mA_n-\frac {1}{1-(AB)^2} B^mB_n,
\end{eqnarray}
\begin{eqnarray}
&& {{\Pi_{(A,B)}}^m}_nA^n\approx 0, \qquad
{{\Pi_{(A,B)}}^m}_nB^n\approx 0, \cr
&& {{\Pi_{(A,B)}}^m}_np^n\approx p^m, \qquad
{{\Pi_{(A,B)}}^m}_n{\Lambda^n}_i\approx {\Lambda^m}_i
\end{eqnarray}
\begin{eqnarray}
&& {{\Pi_{(p,\Lambda)}}^m}_n={\delta^m}_n-
\frac {(p\Lambda_j)\nabla^{ji}p^m\Lambda_{ni}}{(p\Lambda)\nabla(p\Lambda)}
-\frac
{(p\Lambda_j)\nabla^{ji} {\Lambda^m}_i p_n}{(p\Lambda)\nabla(p\Lambda)}+
\cr
&&+\nabla^{ji}{\Lambda^m}_i\Lambda_{nj}- \frac
{(p\Lambda_k)\nabla^{ki} {\Lambda^m}_i (p\Lambda_s)\nabla^{sj}
\Lambda_{nj}}{(p\Lambda)\nabla(p\Lambda)}- \cr
&& -\frac {p^m p^n}{(p\Lambda)\nabla (p\Lambda)},
\end{eqnarray}
\begin{eqnarray}
&& {{\Pi_{(p,\Lambda)}}^m}_n p^n\approx 0, \qquad
{{\Pi_{(p,\Lambda)}}^m}_n {\Lambda^n}_i\approx 0, \cr
&& {{\Pi_{(p,\Lambda)}}^m}_n A^n\approx A^m, \qquad
{{\Pi_{(p,\Lambda)}}^m}_n B^n\approx B^m,
\end{eqnarray}
where $\nabla$ is the inverse matrix to $\Delta$,
$\nabla_{ij}\Delta_{jk}=\delta_{ik}$ and $\approx$ means weak equality.
Note also that $(p\Lambda)\nabla(p\Lambda)\approx\frac 87\neq0$.
In the presence of the projectors the first class constraints can be
written in the form
\begin{equation}
{{\tilde p}_\Lambda}^m\equiv{{\Pi_{(p,\Lambda)}}^m}_n
{{\Pi_{(A,B)}}^n}_k {{p_\Lambda}^k}_i=0.
\end{equation}
At the next
stage, one needs to construct the Dirac bracket associated with all the
second class constraints of the problem, which will look like
\begin{eqnarray}
&&\left\{M,N\right\}_D=\left\{M,N\right\}+
\left\{M,pA\right\}\dots\left\{{p_A}_n,N\right\}+
\left\{M,A^2-1\right\}\dots \cr
&&\left\{{p_A}_n,N\right\}+\left\{M,A \Lambda_i \right\}\dots
\left\{{p_A}_n,N\right\}+\left\{M,pB\right\}\dots\left\{{p_B}_n,N\right\}+\cr
&&\left\{M,B^2-1\right\}\dots\left\{{p_B}_n,N\right\}+
\left\{M,B\Lambda_i \right\}\dots\left\{{p_B}_n,N\right\}-\cr
&&(-1)^{\epsilon(M)\epsilon(N)}(M \leftrightarrow N)+ 
\mbox{\it{terms not involving} $p_A,p_B$},
\end{eqnarray}
where $\dots$ denotes some specific functions and $\left\{M,N\right\}$
is the usual Poisson bracket. As it is seen, under this bracket $A^n,B^m$
commute both with each other and with the first class constraints
${{\tilde p}_\Lambda}^m=0$ from the sector of additional variables.
This implies that the subsequent split of the fermionic constraints
$p_\theta+i\theta\Gamma^np_n=0$ into holomorphic and antiholomorphic
sets will be consistent with the vanishing of the first class
constraints ${{\tilde p}_\Lambda}^m=0$ on physical states. This was the
problem to solve.

Thus, in this letter we have reconsidered the complex
structure in the Gupta-Bleuler quantization scheme, introducing a
gauge fixing procedure based on the addition of a set of auxiliary
variables, which makes the vanishing of the first class constraints on
physical states compatible with the holomorphic-antiholomorphic
splitting of the fermionic constraints. We have built explicitly the
corresponding Lagrangian formulation.

Since this Lagrangian looks like a monster, we have little hope to
be really able to quantize a model on the basis of this scheme.
However, our understanding is that
the Lagrangian above can be viewed as the existence theorem for the complex
structure associated with the $10d$ Casalbuoni-Brink-Schwarz model.

Although the approach
proposed here proved to be too complicated, we expect that the technique will
be efficient when applied to theories possessing a constraint like
$(Ap)=0$, with A {\it a dynamical variable}. One of the possible
applications seems to be the particle in anti-de Sitter space and this
work is in progress now.

\vspace{0.5cm}

{\bf Acknowledgements}\\

One of the authors (A.G.) thanks A.A. Deriglazov and P.M. Lavrov for
useful discussions. His work was supported by INTAS-RFBR grant 95-829 and
by FAPESP.

\end{document}